# Explanation of Lithosphere-Atmosphere-Ionosphere Coupling System Anomalous Geophysical Phenomena on the Basis of the Model of Generation of Electromagnetic Emission Detected Before Earthquake

**M.K. Kachakhidze[1], N.K. Kachakhidze[1], T.D. Kaladze[2]**

[1]*St. Andrew The First-Called Georgian University of The Patriarchy of Georgia, Tbilisi, Georgia*
[2]*Iv. Javakhishvili Tbilisi State University, Institute of Applied Mathematics, Tbilisi, Georgia*

*Correspondence to:* M. K. Kachakhidze    manana_k@hotmail.com

**Abstract**

The present paper considers possible physical mechanisms of the geophysical phenomena which may accompany earthquake preparation process and expose themselves several months, weeks or days prior to earthquakes. Such as:

- Changing of intensity of electro-telluric current in focal area;
- Perturbations of geomagnetic field in forms of irregular pulsations or regular short-period pulsations;
- Perturbations of atmospheric electric field;
- Irregular changing of characteristic parameters of the lower ionosphere (plasma frequency, electron concentration, height of D-layer etc.);
- Irregular perturbations reaching the upper ionosphere, namely F2-layer, for 2-3 days before the earthquake;
- Increased intensity of electromagnetic emission in upper ionosphere in several hours or tenths of minutes before earthquake;
- Lighting before earthquake;
- Infrared radiation;
- Total Electron Content (TEC) anomalies.

Physical mechanisms of mentioned phenomena are explained on the basis of the model of generation of electromagnetic emission detected before earthquake, where a complex process of earthquake preparation and its realization are considered taking into account distributed and conservative systems properties.

Since the above listed Lithisphere-Atmosphere-Ionosphere (LAI) system geophysical phenomena are less informative with the view of earthquake forecasting, it is admissible to consider them as earthquake indicators.

## §1. Changing of Intensity of Electro-Telluric Current in Focal Area And ULF Magnetic Pulsations

A quite large variety of physical mechanisms were proposed for the generation of electromagnetic signals possible earthquake precursors, including electro-kinetic phenomena, effects linked to defects in condensed matter, piezo-electric phenomena, exo-electron emission and etc. (Telesca et al., 2013).

It is proved, that dynamic processes in the earthquake preparation zones can produce current systems of different kinds (Molchanov and Hayakawa, 1998; Kopytenko et al., 2001) which can be local sources for electromagnetic waves at different frequencies, including ULF. High-frequency waves attenuate so rapidly that they cannot be observed on the Earth's surface, whereas ULF waves can propagate through the crust and reach the Earth's surface (Sasai and Ishikawa, 1997; Huang et al., 1999; Kopytenko et al., 2006; Liu et al., 2006; Gu et al., 2006; Chen et al., 2011; Hattori et al., 2004; Varotsos, et al., 2010).

Thus, in ground-based observations, we could expect some ULF signals of seismic origin of the order (0.01 Hz), observed in both geoelectric and geomagnetic fields (Mizutani, et al., 1976; Kopytenko, et al., 2001; Surkov, et al., 2002; Dunson, et al., 2011; Uyeda, 2013).

Furthermore, it should be stated that with the view of reliability studies results of geomagnetic field perturbation it is necessary to exclude magnetosphere influence (Masci, 2011). As for electrotelluric variations before an earthquake that take place in an earthquake preparation zone different attempts have been made to explain the generation of them (Varotsos and Alexopoulos, 1986; Slifkin, 1993; Utada, 1993; Lazarus, 1993; Teisseyre, 1997; DU Ai-Min, et al., 2004). Perturbation of telluric field is considered as a factor of such significance that series of papers (Varotsos and Alexopoulos, 1984 a,b; Varotsos, et al., 1988; Varotsos, et al., 1993) suggested that telluric variations can be used as practical tools for the short-term prediction of earthquakes.

According the computations made by Chapman and Witenhead, meridian telluric component variation is generated by induction as a result of latitudinal magnetic component variation, while meridian magnetic component variation is immediate result of latitudinal telluric component variation (Chapman and Whitehead, 1922).

The above stated shows that during earthquake preparation in the epicentral area, local electric and magnetic fields should suffer variations, and frequencies of electric and magnetic fields should be equal, which was proved experimentally (Hattori, 2004).

In special scientific literature magneto-telluric field is considered as a field that is perturbed by local and regional factors (Kraev, 2007). At the same time at perturbation telluric field assumes vertical direction and becomes linearly (or plainly) polarized (Kraev, 2007). Practically perturbed magneto-telluric field should be revealed itself in earthquake focus, on the very first stage of formation of the main fault; this fact was confirmed by laboratory and field observations (Varotsos, 2005; Molchanov and Hayakawa, 2008; Papadopoulos, et.al., 2010; Orihara, et al., 2012).

Telluric field perturbation and polarization during earthquake preparation are contributed by:
- the processes connected with crack forming in epicentral area;
- asymmetric generation of waves in perturbation sources;
- wave spreading in anisotropic medium;
- wave refraction at the border of two mediums - earth-atmosphere;.

Experimental investigations proved that the earth currents generate instant magnetic variations (Kraev, 2007).

According to our model (Kachakhidze, et al., 2014), due to the fact that in earthquake focal zone a contour is formed, telluric field present in the limits of this contour will play the role of the so-called "displacement current". In addition to it induction vector $\frac{\partial \vec{D}}{\partial t}$ will be directed vertically along $OZ$ axis, from the fault band towards earth surface (Fig.1). Magnetic field lines of forces induced by this field will create the right system at the vector $\frac{\partial \vec{D}}{\partial t}$ which causes the earth magnetic field perturbation because $Y$ component will be directed from east to west.

Induced magnetic field – while spreading in the zone between the earth and ionosphere, will condition magnetic pulsations in this space.

It should be taken into account that at this stage of earthquake preparation there are no effects conditioned by "displacement currents", small perturbations of magneto-telluric field can be expressed only in the increase of the so-called "polarization vector" ($B_z/B_y$), which is proved experimentally too (Hattori, 2004). In case of existence of "displacement current" decrease of the "polarization" parameter ($B_z/B_y$) ratio will take place, since at the growth of telluric current perturbation effect of "displacement current" will overlap effect of natural magneto-telluric field and $B_y$ value of component of magnetic field will increase too, which is observed in nature (Hattori, 2004, Dudkin, et.al., 2013). At the same time, growth of tectonic stress in focal zone results in increase of fault length, which, in its turn, contributes to increase of perturbation area of telluric field.

Due to the fact that perturbed telluric field acquires vertical direction to the earth surface, it can play a role of an antenna for ULF signal.

It should be taken into consideration that magneto-telluric field perturbation will take place not only during period preceding earthquake, but also after it too, till tectonic stress accumulated in focal area is released completely (Moroz, et al., 2004).

### §2. Perturbations of Atmospheric Electric Field

It is known from scientific literature that thanks to contact of solid or gaseous phases existing between two mediums, the earth and atmosphere, diffusion of electrons and ions and ion adsorption take place, which conditions creation of stable electric layer (dipole layer) on the contact. In this layer the electric field, supported by factors conditioned by earthquake preparation are accumulated, which can be called "additional" electric field and can be marked as $E_n$ (Kraev, 2007. In this case, electric field potential at the separating border of these two mediums suffers discontinuation, which equals to contact "additional" electric field strength:

$$\varepsilon^{add} = \int_1^2 E_n^{add} dn$$

where 1 and 2 points are located on both near coasts of contact surface. It is clear that the fact of mentioned field discontinuation will be expressed on all geophysical phenomena connected with "additional" field.

It should be emphasized that atmospheric electric field potential gradient in the period of earthquake preparation suffers changes not only at the influence of earth currents but also at the influence of atmospheric parameters. Therefore, only the so-called "filtered" significance of electric field potential gradient of atmosphere should be considered as the "additional" field (when the influences of meteorological parameters and all types of fluctuations are excluded) (Kachakhidze, et.al., 2009).

It is known from scientific literature that vertical electric current of atmosphere is connected with oscillations of elements characteristic to the earth magnetic field. Alongside with it, earth electric current is connected with the electric current, which is directed from atmosphere to the earth. Thus, vertical current oscillastions, and correspondingly, gradient almost always are accompanied by the earth current vibrations (Chijevski, 2007).

It is proved counter-connection among variations of atmospheric electric and telluric fields. For the fixed time moment we can write:

$$\mathcal{E}(t_0) = E_{tellur.} \times l + E_{atmosp.} \times l_a$$

where $l$ and $l_a$ are equivalent lengths relevant to telluric and atmospheric electric fields; the higher $E_{tell.}$ the lower is $E_{atm}$ and vice versa if $l$ and $l_a$ are of the same order (Kraev, 2007)..

According to Wilson's theory earth and ionosphere play the role of the capacitor plate. Alongside with it according to Erenkel's theory atmospheric electric field is completely explained by the processes going on in troposphere, by polarization of clouds and their connection with the earth (Chijevski, 2007).

It should be emphasized that during "fine" weather the global factors play preferential role, while during "bad" weather - local ones, since unitary variations, generally are connected with variations of total electric charge of the earth, while those of local – are connected with variations of values of volumetric electric charges in atmosphere of definite region and distributions by height.

In case of "fine" weather the earth surface potential is negative, potential of the earth surface boundary layer of atmosphere – positive, potential of the top of atmosphere – negative and that of the lower layer of ionosphere –positive.

As to the period of earthquake preparation because of presence of "additional" field, the picture of the earth-ionosphere system electric strength stress distribution is changed: if potential gradient of "additional" field" will exceed electric field potential gradient of atmosphere, electric field strength between the earth and ionosphere will be directed from the earth surface to ionosphere, that is the earth surface potential will be positive, while that of ionosphere – negative. In this case the earth surface boundary layer of atmosphere will have negative potential, while the top of atmosphere, near the ionosphere – positive.

This inverse state of atmospheric electric field, on the one hand suffers pulsation by frequency inherent to polarized perturbed telluric field strength conditioning it, and on the other hand it is affected by electromagnetic field caused by earthquake preparation. As a result of superimposition of these fields we receive variation of atmospheric electric field potential gradient in rather big ranges (in case of "non- filtered" field), which is proved by experiments too (Smirnov, 2008; Kachakhidze, et.al., 2009; Pulinets, 2009; Harrison, et al., 2010; Silva, et al., 2012). Atmospheric electric field will stay inversed, till telluric current perturbations disappear and it will go back to its background data limits.

In case of large earthquakes this process continues several days and during this time "additional" field plays the role of a "supporting force" with the view of preservation and strengthening of atmospheric inversed electric field. Reduction and gradual disappearance of this field will take place only when tectonic stress in earthquake focus is released completely, that is, when a series of aftershocks will end.

When perturbed telluric current causes atmospheric electric field inversion, the principle of global electric contour closure is destroyed and in the global electric circuit system a defined zone of anomalous electric field is created (approximately above earthquake focus). In this case lines of force of atmospheric electric current are no more closed, that is, for definite time, it, passing to atmosphere, doesn't leave the limits of the earth-ionosphere and accumulates charges only at the ends of lines of force. Simultaneously at this period "displacement current" in atmosphere exceeds conduction current and it governs all atmospheric processes.

Thus, it turned out that a picture of atmospheric electric field strength during earthquake preparation period, partially repeats a picture of the so-called "bad weather" of electric field strength.

If we take into consideration recent works, according to which in the period of earthquake preparation magnetosphere suffers perturbation and from there high energy particles, namely electrons invade atmosphere (Koldashov, et al., 2013), it is not excluded that the terms for generation of clouds are created there. It should also be underlined that at this moment there is short and long-wave electromagnetic emission in atmosphere, within radiodiapason, emitted from the earth surface.

Due to the fact that a picture of electromagnetic field distribution in earthquake preparation period coincides with a picture of distribution of "bad" weathe fields, it can be admitted that the above described factors are the causes of existence of clouds during large earthquakes which are mostly fixed in the vicinity of earthquake epicentral zone, which by majority of scientists is attributed justly to local perturbation conditioned by earthquake preparation process (Guo and Wang, 2008.; Genzano, et al., 2009; Guangmeng. and Jie, 2013; Harrison, 2004; Harrison, et al., 2009; Harrison, et. Al., 2010).

## §3. TEC anomaly

Recent experimental studies revealed interrelation between particles ejection from the Earth radiation belt and geophysical processes: seismicity, thunder phenomenon etc (Boyarchuk, et. Al., 2007; Koldashov, et al., 2013).

It was also proved that the ultralow frequency component (below 10 Hz) of electromagnetic radiation spreads in magnetosphere and reaches lower limits of radiation belt (RB), almost not absorbed in atmosphere and ionosphere. Between the high energy charged particles (electrons and protons with energies of the order of some dozens MeV) of radiation belt and ULF radiation there is a quasi resonance type interaction. As a result, particle ejection takes place at the height below the atmospheric border of the radiation belt (Boyarchuk, et al., 2007; Koldashov, et al., 2013).

According to the same works, first: the higher proton and electrone energy, the closer their capture zone to the Earth. There is also the second, very important condition: particles might be present in the capture zone, if the lower limits of the zone passes above the upper limit of the atmosphere. Otherwise, particles, travelling in the capture zone can fall in the dense layers of the atmosphere, and worse than that – on the Earth, where they perish.

Exposure of stable belt of electrons with high energy belongs to the researchers of National Research Nuclear University MEPhI (Moscow, Russia) (Boyarchuk, et. al., 2007; Koldashov, et al., 2013).

It is known that several days before large earthquakes in the lower ionosphere an abnormal TEC variation over the epicenter is confirmed (Boudjada, et.,al., 2013; Hayakawa, 1999; Hayakawa, et al.,2013; Namgaladze, et., al., 2012; Pulinets, 2012; Ouzounov, et al., 2013).

According to our model in the earthquake preparing period, because of atmospheric electric filed inversion a top of atmosphere has positive potential (§ 2), which should contribute to trapping of emitted electrons. At the same time, at this moment ionosphere will have lower potential compared to that of the Earth surface.

High energy particles are classified according to characteristic energies. Relatively low energy particles (mainly electrons) will deposit in the zone above perturbed (inversed) atmospheric electric field, the lower boundary of which passes above upper boundary of atmosphere, which will contribute to the formation of TEC anomaly section in ionosphere above earthquake focus (if we don't take into account polarization conditioned by "additional" field).

This process will be kept till atmospheric electric field inversion takes place, that is till the top of atmosphere will have positive potential, although at this moment there should exist ULF magnetic field, bottom-up directed (from the Earth surface to ionosphere) the existence of which is observed experimentally in earthquake preparation period.

Results of ARINA and VSPLESK satellite experiments, presented by Russian sciences(Koldashov, et al., 2013), show that, along with bursts of particles observed at various longitudes of disturbed L-shell, there are bursts of particles, grouped directly in the regions of local disturbances of the radiation zone (seismic and lightning activity zones).

The results of observations of dynamics of flux of high-energy electrons in the near-Earth space during the development of seismic event in Japan in March 2011, once again confirm the existence of seismo-magnetosphere interrelation and demonstrate the possibility of using this phenomenon for satellite monitoring of earthquakes (Koldashov, et al., 2013).

Considering the above stated we can make the following conclusion: ULF magnetic field, perturbs magnetosphere from where high energy particles are ejected (mainly electrons) (Boyarchuk, et. al., 2007; Koldashov, et al., 2013).

According to our model, at this moment atmospheric electric filed suffers inversion, that is the Earth possesses positive potential, while the ionosphere – negative and below ionosphere, top of atmosphere – will have positive potential. It is also important that according to observations, atmospheric electric field potential gradient values vary within a rather wide diapason and have pulsating character. Thus, e.g. in case of the Caucasus $M \geq 5$ earthquakes, maximum and minimum values of gradients of atmospheric "unfiltered" electric field potential in various cases vary within (3970 V/m - (-3500) V/m) (Kachakhidze, et., al., 2009; Silva, et.al., 2012).

Due to the fact that electric energy in electromagnetic wave at any moment is equal to that of magnetic, ULF field can play a role of unipolar antenna too. It is clear that atmospheric pulsating electric field can play a role of magnetosphere perturbation factor which will be intensified some time before earthquake occurrence, since variation of atmospheric electric field potential gradient reaches its maximum namely in this period.

In such situation, definite part of high energy electrons ejected from perturbed magnetosphere, bursting in ionosphere will condition its perturbation and if its energy will not be high enough to overcome top of atmosphere, where it is met with relatively high density medium, this part of electrons will be deposited in relatively low strata of ionosphere, boarding the top of atmosphere.

This fact is contributed by the situation that at this moment top of atmosphere has positive potential. Therefore it holds the lower edge of the ionosphere and keeps it in a balanced state, where there already exists TEC anomaly. This situation will be retained till there is atmospheric electric field inversion and magnetosphere perturbations induced by ULF magnetic field. This group of high energy particles can be considered as the "first group" electrons.

Particles ejected from magnetosphere (mainly electrons), which possess higher energies than the electrons of the above stated "first group" will overcome top of atmosphere boundary and will burst in atmosphere. At the interaction of accelerated electrons with atmospheric ones the «red sprites», "bluejets", "elves" and gamma-quant flux will be formed, which in satellite investigations are detected as gamma – bursts (Boyarchuk, et. al., 2007; Koldashov, et al., 2013). Definite quantity of these particles (conditionally particles of the "second group") of course perish, in the top of atmosphere, which should be accompanied by release of heat.

It may be, that outgoing long wavelength radiation (OLR) on the top of the atmosphere, fixed by experiments, is conditioned namely by this effect (Liu, 2000; Ouzounov, et al., 2013).

Particles of the so-called "third group" distinguished from those of the first two groups, which possess far more energy than particles of the "first" and the "second" groups, bursting in atmosphere, continues its way towards the positive potential Earth surface, but perish at the contact with the Earth surface and definite heat release takes place. Presumably positive temperature gradient that is fixed on the earth surface is conditioned namely by it this reason. Of course, other thermal effects of the processes going on in the earthquake focus should be taken into account (Dunajecka, et.,al., 2005; Tramutoli, et. al., 2005; Pulinets, et., al., 2006; Genzano,et.,al., 2009; Saradjian, et., al., 2011).

Magnetosphere perturbations cause by ULF magnetic field of course will result in :
Irregular changing of characteristic parameters of the lower ionosphere (plasma frequency, electron concentration, height of D-layer etc.); Irregular perturbations of the upper ionosphere, namely F2-layer, for 2-3 days before the earthquake; Increased intensity of electromagnetic emission in upper ionosphere in several hours or tenths of minutes before earthquake.

According to a recent statistical analysis of the DEMETER data (Stangl, et., al., 2011; Zhang, et., al., 2013) of the ion density of the ionosphere, it was found that there are more perturbations for earthquakes with their epicenter below the sea and also that the intensity of perturbations is more enhanced for sea earthquakes than for inland earthquakes. Alongside with it, according to observations, the ionospheric perturbation is seen mainly over the sea (Hayakawa, et., al., 2013).

And really, because of high electric conductivity of ocean water, effect of the perturbed telluric field should be strengthened. In such conditions it is admissible to transmit radio-frequency electromagnetic radiation into the ionosphere, the might take place with the electrolytes in the ocean acting as an antenna (Daniel S. Helman, 2013).

If the focus of the incoming earthquake is located under water layer, of course alongside with generation of ULF diapason waves generation of VLF electromagnetic emission should take place. Although in case of sufficient depth, water layer will absorb VLF electromagnetic emission and will let only ULF diapason waves pass.

If telluric current contour will consists of section of coastal band or shallow waters, earth electromagnetic emission spectrum should reveal itself fully.

### § 4. Infrared Radiation

There is a theoretical model, that communicates with each other 0.7-20 micrometer wavelength atmospheric IR-radiation intensification in epicentral area of incoming earthquake and anomalous variation of atmospheric electric field value (Meister, et al., 2011; Liperovsky, et al., 2008). This phenomenon in the cloud is interpreted on the basis of Frankel theory (Frenkel,1949) and is developed on the background of electric field direction inversion.

This theoretical model requires specific terms, in particular, necessity of sufficient density of atmospheric aerosol in incoming earthquake epicentral area and emanation of radon from the Earth as a source of alpha-particles. Quasi-stationary action of this factor should cause anomalously strong pulsations, that is spikes, of local atmospheric electric field, generation process of which should continue for a rather long time (1-100 minutes). The above referred work shows that in electric field of anomalously great strength, atmospheric charged particles can be accelerated to a rate that their energy reachs the limit needed for generation of IR radiation.

A model (Liperovsky, et al., 2008) is interesting with the view of interpretation of a possible mechanism of IR radiation generation in lower atmosphere, but its realization, besides rather intense radon emanation requires satisfaction of rather strict conditions. Namely spike amplitude must reach anomalously big value for atmospheric electric field strength (from 1000 to 3000 V/m).

Alongside with it, product of length of free run of charged particles ($CO_2$ and $CH_4$) in the clouds and electric field strength and charge value determines particle energy, which is necessary for generation of 2-15 mcm wavelength IR radiation. The paper Liperovsky et al., 2008, shows that in case of anomalously strong electric field for generation of IR radiation in normal height clouds, it is enough that minimal free run length should be ~ 7 mcm.

It is natural that together with the growth of height of clouds, free run length will increase because of decrease of medium density. Therefore, infrared radiation might be caused by relatively less intense atmospheric electric field spike.

But if we rely on the above describe (§3) experimental observations which prove the bursting in of high energy particles to the Earth atmosphere in the period before earthquake because of magnetosphere perturbations caused by ULF magnetic field, the anomalous oscillations in wide diapason (3970 V/m - (-3500) V/m) of atmospheric electric field strength (Kachakhidze, et., al., 2009; Silva, et., al., 2012 and the presence of atmospheric electric field inversion according to our model (§**2**), all restrictions referred to above will be removed and the possibility of generation of infrared radiation in atmosphere will be clearly explained.

## §5. Lighting

In scientific literature are fixed the experimental facts of earthquake lights — mysterious glows sometimes reported before or during seismic shaking — finds that they happen most often in geological rift environments, where the ground is pulling apart (Bagnaia, et al., 1992; St-Laurent, et al. 2006; Atzori, et al.,2009; Stephan, et al., 2009; Fidani, et al., 2010).

As it was stated above, there are investigations based on experimental data according to which high energy particles bursting in atmosphere takes place during development of the seismic event (Akhoondzadeh, et al., 2010; Ouzounov, et al., 2011; Zhang, et al., 2013; Koldashov, et al., 2013).

According to our model (§2), in the area of earthquake preparation, atmospheric electric field suffers inversion. Besides, within several days preceding the earthquake (10-15) atmospheric electric field potential gradient values undergo variation in rather wide diapason (3970 V/m - (-3500) V/m) (Kachakhidze, et al., 2009; Silva, et al., 2012) as a result of which between the top of the atmosphere and the ionosphere non-stationary electric field is created.

According to the woks of Russian scientists, at the terms of magnetosphere perturbed by ULF magnetic field emission, electrons acceleration up to energy of dozen order MeV takes place. Then electrons interact with the upper atmosphere, ionize it and avalanche-like generation of electrons takes place, leading finally to electric discharge. As is known, at the interaction of accelerated particles with atmosphere, the conditions are created for lighting. Lighting is emitted not by particles as such, but by chemical element atoms incited by them, which exist in the upper atmosphere (Galper, et al., 2006; Boyarchuk, et al., 2007; Koldashov, et al., 2013).

Thus, based on a model offered by us and experimental observations of magnetosphere perturbation conditioned by ULF magnetic field the causes of lighting generation that is observed before earthquakes are explained fully.

The above referred study, which relies upon electrodynamics enables us to make the following conclusion: electromagnetic emission detected in the period preceding earthquake is a main precursor and in LAI system it results in various anomalous geophysical phenomena.

**Conclusions**

The physical principles of generation of anomalous geophysical phenomena in LAI coupling system, accompanying earthquake development process, are explained on the basis of analogous model of electromagnetic contour and classical electrodynamics.

- Seismogenic area belongs to oscillation system in the process of earthquake preparation;

- Focal area from the moment of earthquake preparation starting till the ending of aftersoks series (including foreshoks and main shock) combines properties of two systems – distributed and conservative;

- From the starting moment of avalanche-like crack formation, processes going on in the earthquake focal zone (as conservative system), can be explained by classic electrodynamics;

- After earthquake occurrence (and after attenuating of aftershocks) focal zone will bear only distributed system signs and properties;

- In earthquake preparation period in the earthquake focus a contour is formed which emit VLF/LF frequency electromagnetic waves;

- In earthquake preparation period processes going on in earthquake focus cause electro-telluric field perturbation in focal zone, which, in its turn, is a reason of generation of ULF-magnetic field pulsations;

- Strength of perturbed electro-telluric field results in atmospheric electric field inversion; besides, in this period in the vicinity of epicentral zone, the Earth surface has positive potential;

- In epicentral zone type of atmospheric electric field potential gradient variation coincides with the "bad" weather situation, at a definite accuracy; clouds are formed which are stimulated by the presence of high energy particles in this zone;

- Part of long wavelength electromagnetic field emitted from apicentral zone - anomalous ULF magnetic field causes magnetospheric perturbation, because of which high energy particles, mainly electrons bursting in ionosphere, deposit on lower boundary of the ionosphere and result in electron and electron density variation in the ionosphere (GPS/TEC);

- Particles of relatively high energy continue their way to the Earth, but their one part fails to pass through relatively dense medium of atmosphere and perishes on the top of the atmosphere; this might be a reason of outgoing long-wavelength radiation (OLR) on the top of atmosphere;

- Particles with very high energy bursting in the atmosphere, reach the Earth and at the contact with the Earth surface perish. This can result in insignificant increase of temperature on definite areas of the Earth surface;

- As to lighting and infrared radiation here the known physics works, taking into account that high energy particles bursting in from perturbed magnetosphere to atmosphere, where atmospheric electric field inversion and anomalous oscillations of values of atmospheric electric field strength in wide diapason take place;

The earthquake preparation process causing various anomalous physical phenomena, with the scientific point of view is a rather complex subject of multi-branch study, and because of it, it needs wide-scale theoretical and experimental works in future;

The authors of the present paper consider that electromagnetic emission fixed before earthquake, which, apparently offers us relatively thorough information enabling us to make prognostic conclusions, should be considered as a "main precursor" of the earthquake, or simply, as a "precursor" (Kachakhidze, et al., 2014. arXiv. 1407.3488), while the phenomena which are fixed in the process of earthquake preparation, having direct connection with processes in the focal area, but don't enable us to make diagnostic conclusions, can be considered as "earthquake indicators".

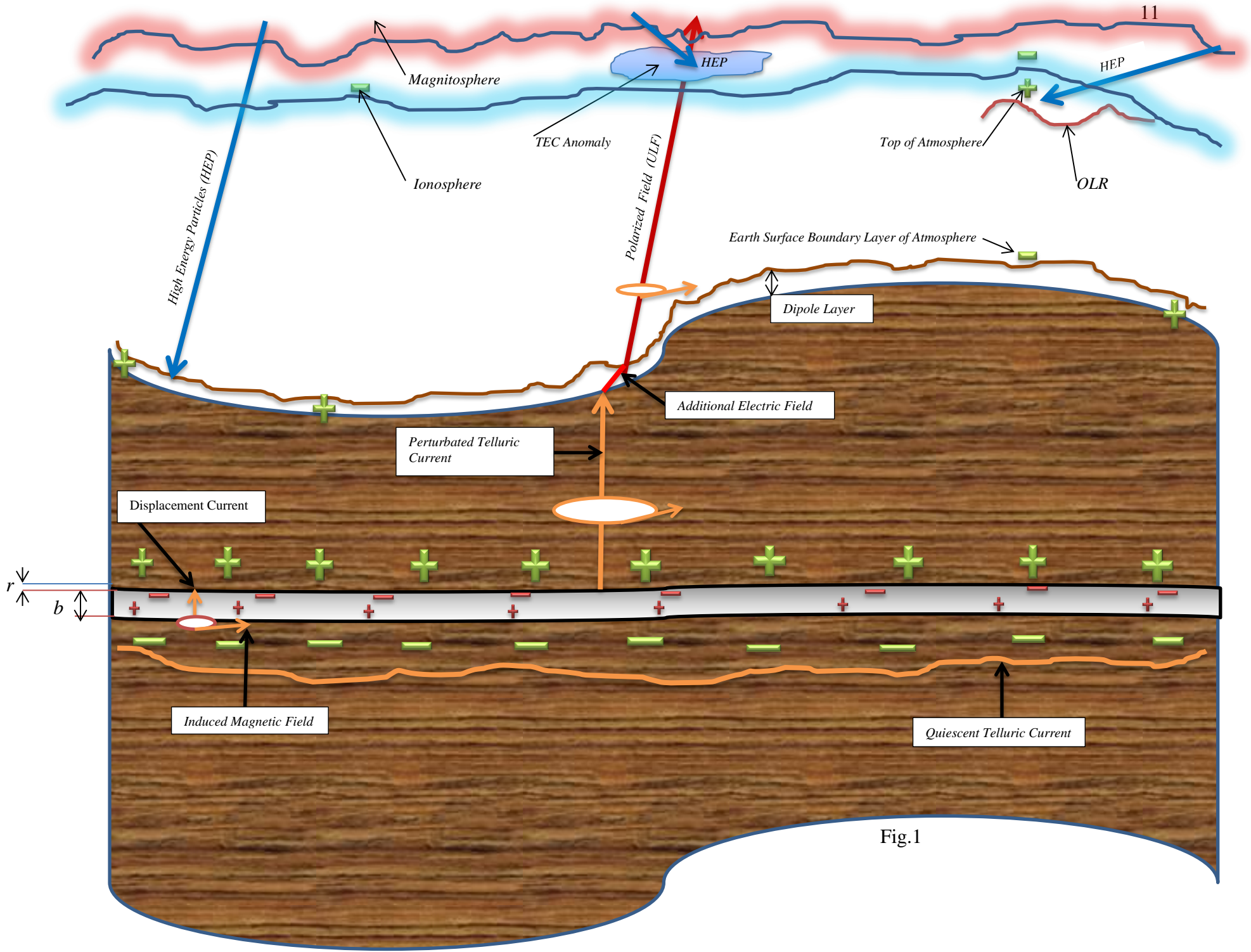

Fig.1